\documentclass[prl,aps,twocolumn,showpacs,amsmath,amssymb,floatfix,citeautoscript]{revtex4-1}

\usepackage{graphicx}
\usepackage{dcolumn}
\usepackage{amsmath}
\usepackage{amssymb}
\usepackage{times}
\usepackage{epsfig}

\begin{document}

\title{Lifshitz transition in Kondo alloys }

\author{S\'{e}bastien Burdin}
\affiliation{Univ. Bordeaux, LOMA, UMR 5798, F-33400 Talence, France}
\affiliation{CNRS, LOMA, UMR 5798, F-33400 Talence, France}
\author{Claudine Lacroix}
\affiliation{Institut N\'eel, Centre National de la Recherche Scientifique (CNRS) \& Universit\'e Joseph Fourier (UJF),
25 Avenue des Martyrs, F-38042 BP166 Grenoble Cedex 9 France}

\begin{abstract}
We study the low energy states of Kondo alloys as function of the magnetic impurity concentration per site, $x$, and the conduction electron average site occupation, $n_c$. Using two complementary approaches, 
the mean-field coherent potential approximation and the strong coupling limit, we identify and characterize two different Fermi liquid regimes. We propose that both regimes are separated by a Lifshitz transition at $x=n_c$. Indeed, we predict a discontinuity of the number of quasiparticles which are enclosed in the Fermi surface. 
This feature could provide a scenario for the non-Fermi liquid properties that were recently observed in Kondo alloy systems around $x=n_c$. 
\end{abstract}

\date{\today}

\maketitle

Kondo alloys are realized in broad families of strongly correlated
materials where magnetic quantum impurities are embedded randomly in a crystalline system with conduction
electrons~\cite{Hewsonbook1993}.
The Kondo problem has been intensively studied for almost half a century~\cite{Hewsonbook1993}, and
the single impurity model was solved exactly by various 
methods~\cite{Wilson1975, Bulla2008, Andrei1980, Wiegmann1980, Affleck1991}. 
In the 70's, Nozi\`eres has adapted Landau's approach~\cite{Landau1957, PinesNozieresbook1966},
showing that a single Kondo impurity can be described universally at low energy as a local Fermi
liquid (LFL)~\cite{Nozieres1974}.
Multiple-impurity models have later been introduced in order to describe systems with a high concentration of Kondo ions. It appeared that, unlike the single impurity models, the multiple-impurity models are not belonging to a single universality class, but can rather give rise to a large variety of ground states
depending on the lattice structure, the electronic occupation, and on the strength of the Kondo coupling which can be tuned with pressure~\cite{Doniach1977,Lacroix1979} .

Putting aside the relevant issues of ordering, we concentrate here on the coherent Fermi liquid (CFL) ground state that is 
observed at low temperature in lots of dense Kondo systems~\cite{Flouquet2005}.
The electronic exhaustion problem~\cite{Nozieres1985,Nozieres1998}  suggested by Nozi\`eres remained a long standing issue for years~\cite{Lacroix1985, Lacroix1986, Tahvildar1997,Tahvildar1998,Tahvildar1999,
Coqblin2000,Burdin2000,Costi2002,Coqblin2003} and was finally understood~\cite{Burdin2000,Nozieres2005}: the Kondo temperature $T_K$ that characterizes the crossover to the low temperature Kondo screening regime, and the Fermi liquid energy scale $T_{coh}$ that characterizes the CFL are indeed the same energy scale. It was shown that the ratio $T_K/T_{coh}$ still depends on the electronic filling and lattice
structure~\cite{Burdin2000,Burdin2007,Burdin2009}.
But this ratio was proven not to depend on the Kondo coupling as it was wrongly thought
initially~\cite{Burdin2000, Costi2002, Nozieres2005}. 

A simple picture of the periodic dense Kondo lattice model had been provided within the strong Kondo coupling
limit~\cite{Lacroix1985, Nozieres1998, Oitmaa2003, Kim2007, Kaul2007}.
Even if this limit does not correspond to the experimental reality, its qualitative validity
is supported by poor man's scaling analysis~\cite{Hewsonbook1993,Anderson1970},
which shows that the low temperature Kondo physics of the single impurity model 
is renormalized to a strong coupling fixed point.
In the infinite coupling limit, the ground state of a Kondo lattice made of $N$ sites with
$N_S=N$ spins $1/2$ is characterized by the formation of
$N_c$ local Kondo singlets, where $N_c$ denotes the number of conduction electrons. The CFL is recovered
from a perturbation expansion at lowest order in $hopping/coupling$.
The corresponding fermionic quasiparticles correspond to the bachelor Kondo spins, whose double occupancy is forbidden, giving rise to strong correlation effects. This strong coupling description also provides the correct number of quasiparticles: $N-N_c$ free spins, which, according to a particle-hole transformation, give $2N-(N-N_c)=N_c+N$
quasiparticles (the factor $2$ comes from the spin-degeneracy).
This result is in perfect agreement with Luttinger theorem~\cite{Abrikosovbook1963},
which predicts an enlargement of the Fermi surface due to the contribution of the 
Kondo impurities~\cite{Nozieres1998,Burdin2000}.

Our analysis starts here: on the one side the universal single impurity Kondo model can be described as a LFL. This picture describes well Kondo systems with very dilute impurities.
On the other side, many realizations of dense Kondo systems with a periodic lattice of magnetic ions have a CFL ground state.
Can these two Fermi liquids be connected continuously to each other at zero temperature?

We consider a Kondo alloy model (KAM), defined by the following Hamiltonian
\begin{eqnarray}
H=\sum_{ij\alpha}t_{ij}c_{i\alpha}^{\dagger}c_{j\alpha}
+J_{K}\sum_{i\in K}{\bf S}_{i}\cdot {\boldsymbol{\sigma}}_{i}~,
\end{eqnarray}
where $c_{i\alpha}^{(\dagger)}$ denotes annihilation (creation) operator of
a conduction electron with spin $\alpha=\uparrow,\downarrow$ on site
$i$ of a periodic lattice that contains $N$ sites. $t_{ij}$ denotes intersite hopping energy, and $J_{K}$
is a local Kondo antiferromagnetic interaction between local quantum spin $1/2$
denoted ${\bf S}_{i}$ and the local density of spin of conduction electrons,
${\boldsymbol{\sigma}}_{i}$.
These $N_S\equiv xN$ Kondo spins, with concentration $x\le 1$,
are located on a subpart $K$ of the periodic lattice. We choose a frozen configuration for $K-$sites, whose positions are distributed randomly without spatial correlation.
The number of conduction electrons is fixed to $N_c\equiv n_c N$, and we restrict $n_c\le 1$ for particle-hole symmetry reason.

\begin{figure}[hhh]
\includegraphics[width=6.6cm, origin=bc, angle=-90]{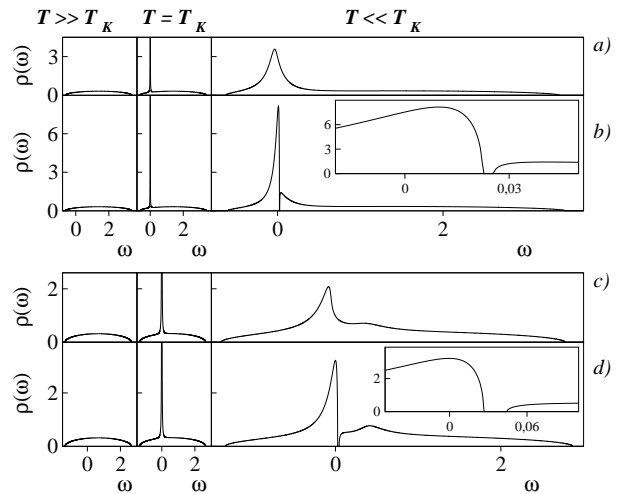}
\caption{Local d.o.s. on a Kondo site for $n_c=0.20$ ($a$ and $b$) and $n_c=0.60$ ($c$ and $d$), and impurity concentrations $a)$ $x=0.10$, $b)$ $x=0.30$, $c)$ $x=0.50$, and
$d)$ $x=0.70$. Left column: elliptic d.o.s. obtained at $T\gg T_K$. Second column: formation of the Kondo resonance, at $T\approx T_K$. Right column: comparison between the dilute LFL ($a$ and $c$) and dense CFL ($b$ and $d$) regimes.
Insets: hybridization gaps above the Fermi level, which characterize the CFL regime.
Parameters: $J_K=1.5$ and electronic bandwidth $W=4$, with $T_K=0.099$ ($a$ and $b$) and 
$0.18$ ($c$ and $d$). 
}
\label{FiguredosKondoalloy_plusinset}
\end{figure}

First, the KAM is studied using the mean-field approximation
for the Kondo interaction~\cite{Lacroix1979,Coleman1983,Read1984}
and considering the random position of $K-$sites within a CPA-DMFT
method~\cite{Burdin2007}, i.e., a matrix version of Dynamical Mean-Field Theory
(DMFT)~\cite{George1996, Metzner1989} which is equivalent to the matrix version of the Coherent Potential
Approximation (CPA)~\cite{Blackman1971, Esterling1975}.
Kondo alloy models had been studied earlier using CPA methods, but focusing mostly on the high concentration of
impurities~\cite{Hoshino1979, Kurata1979,Li1994}. 
In Ref.~\cite{Grenzebach2008} an Anderson alloy model was also studied using CPA and DMFT methods, but with 
a choice of parameters which enforces the system to be always in the CFL regime. 
The matrix CPA-DMFT formalism which
 is presented in details in Ref~\cite{Burdin2007} treats separately Kondo and non-Kondo sites.
For the sake of simplicity the numerical calculations were
performed assuming the conduction electrons move on a Bethe lattice, characterized by an
elliptic non-interacting local density of states (d.o.s.), as depicted on
figure~\ref{FiguredosKondoalloy_plusinset}.
The local d.o.s. on a $K-$site is analyzed as functions of $x$, $n_c$, and the temperature, $T$. At high $T$ (left column on figure~\ref{FiguredosKondoalloy_plusinset}) it corresponds to the non-interacting conduction band, and for $T\approx T_K$ (second column) a Kondo resonance forms around the Fermi level $\omega=0$.
The position of the resonance depends on $n_c$, but its shape at $T\approx T_K$ does not depend on $x$. This is fully consistent with the interpretation that $T_K$ characterizes the incoherent local Kondo singlet formation.
At lower temperature (right column), when a Fermi liquid regime sets in, the local d.o.s.
can exhibit two qualitatively different features: for $x<n_c$
(figures~\ref{FiguredosKondoalloy_plusinset}$a$ and \ref{FiguredosKondoalloy_plusinset}$c$),
it remains roughly similar to the non-interacting one, with just a small deformation around the Fermi level, due to the Kondo resonance.
But for $x>n_c$ (figures~\ref{FiguredosKondoalloy_plusinset}$b$ and \ref{FiguredosKondoalloy_plusinset}$d$)
the Kondo resonance is split by an hybridization gap.
A similar reduction of the d.o.s. had already been suggested
in~\cite{Hoshino1979, Zhang2002}.
This splitting can be explained by the standard two-band picture which is commonly used for Kondo lattices. A systematic analysis of the d.o.s. at low $T$ for different impurity doping and electronic filling clearly shows that the crossover between CFL (with a gap) and LFL (without gap) regimes occurs when $x=n_c$. The dense system remains
Fermi liquid since the gap occurs at $\omega>0$, i.e., above the Fermi level.
However, this mean-field result suggests that the Kondo-interacting (at low $T$) and the Kondo-decoupled (at high $T$)
Fermi liquids can be smoothly connected to each other only for $x<n_c$. 
At larger impurity concentrations (or, equivalently, at lower electronic filling),
a singularity may disconnect the two Fermi liquid regimes, related to this gap
opening.

Studying the strong coupling limit, $J_K\gg\vert t_{ij}\vert$,  provides
a clear evidence of the analytical disconnection between the CFL ($N_S>N_c$) and the LFL ($N_S<N_c$).
Hereafter we generalize to the KAM the exact strong coupling treatment of Ref.~\cite{Lacroix1985}.
\begin{figure}[hhh]
a)~~~
\includegraphics[width=1.1cm, origin=bc, angle=90]{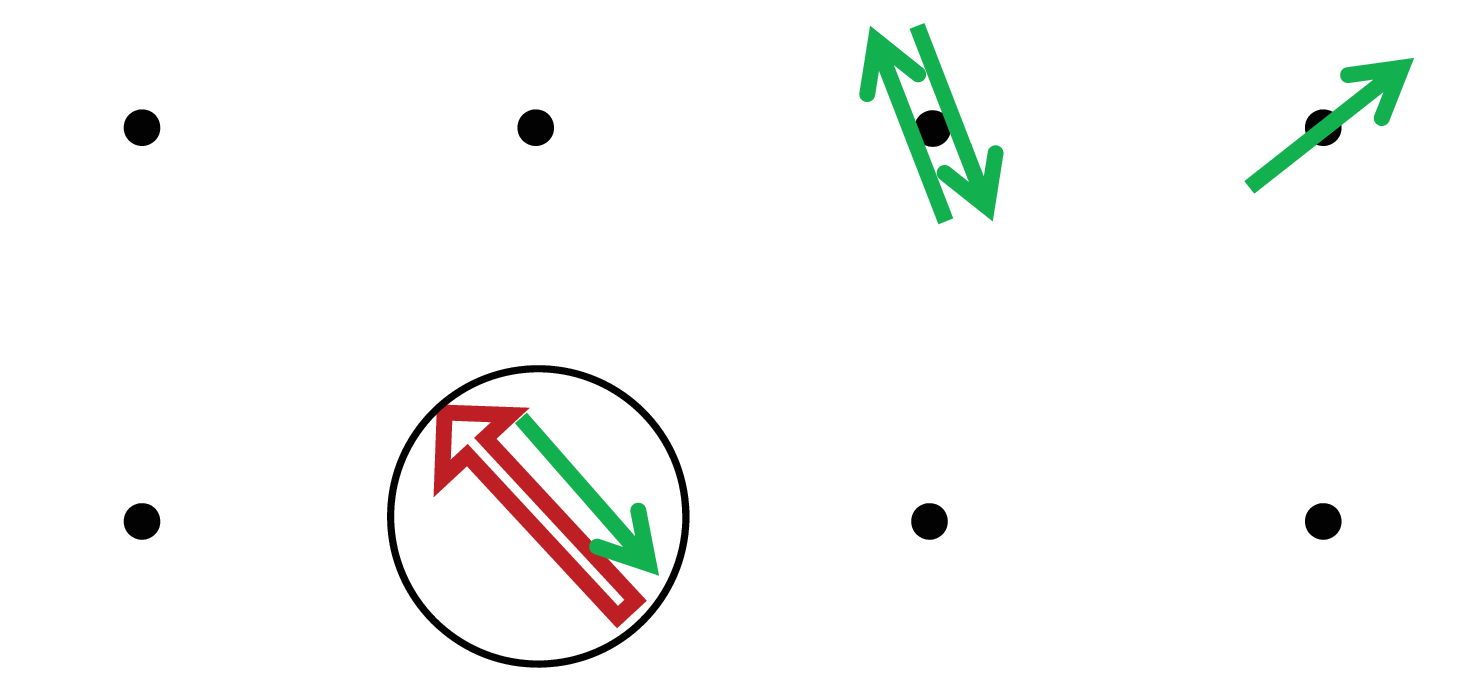}
~~~~~~~~~~~~~~~~~
b)~~~
\includegraphics[width=1.1cm, origin=bc, angle=90]{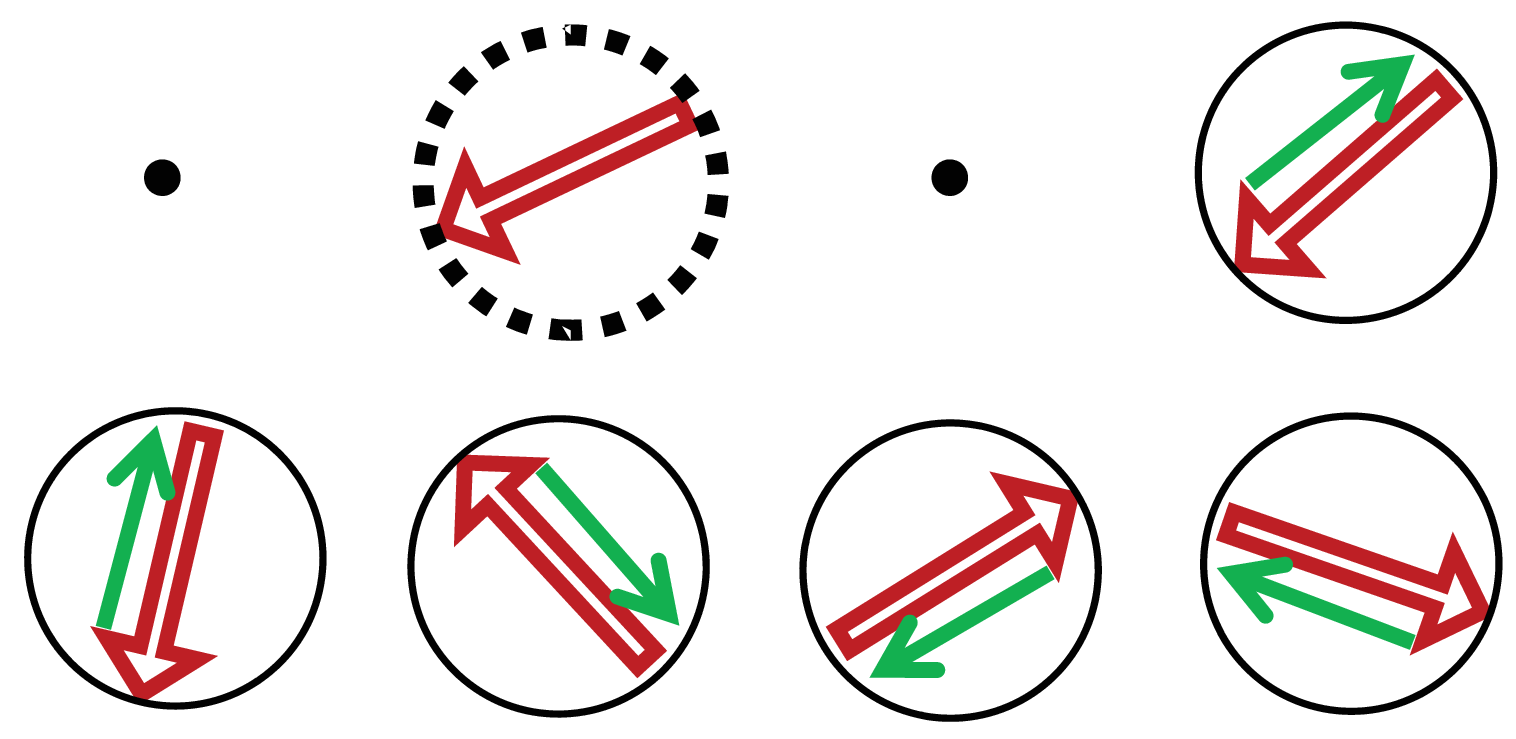}
\caption{Schematic picture of the ground state of the KAM for large $J_K$. 
Kondo spins are depicted by red double arrows, electrons are green simple arrows. 
Black points refer to non-Kondo sites. Kondo singlets are marked by a solid circle.
a) Dilute case. b) Dense case : doted circles indicate unscreened Kondo spins (see text).
}
\label{Figureschemareseau}
\end{figure}
We start with the extreme limit $t_{ij}=0$.
Depending on $N_c/N_S=n_c/x$, we can distinguish two different situations, as illustrated on figure~\ref{Figureschemareseau}: 
For $N_S<N_c$ the ground state is characterized by $N_S$ local Kondo singlets on each Kondo-site.
$N_S$ electrons are thus localized on the
$K-$sites which effectively become "forbidden sites" for the remaining
$N_c-N_S$ electrons. The latter can freely occupy the left $N-N_S$
non-Kondo sites.
This degenerate ground state has a finite entropy per site 
${\cal S}_{LFL}=(1-x)\ln(1-x)-(n_c-x)\ln(n_c-x)-(1-n_c)\ln(1-n_c)$. 
For $N_S>N_c$ the ground state is characterized by the formation of $N_c$ Kondo singlets. The non-Kondo sites are empty, but
a degeneracy results from the various energetically equivalent possible positions of the Kondo screened impurities on $K-$sites.
The resulting entropy per site is
${\cal S}_{CFL}=x\ln(x)-n_c\ln(n_c)-(x-n_c)\ln(x-n_c)$. 
Both ${\cal S}_{LFL}$ and ${\cal S}_{CFL}$ are analytical functions of $x$ and $n_c$, but they cannot be analytically connected to each other. Indeed, both vanish like
${\cal S}\approx -\vert x-n_c\vert\ln\vert x-n_c\vert$ when $x\to n_c$.
The presence of an absolute value in this expression is clearly identified as a
singularity separating the two regimes.
Furthermore, the vanishing of ${\cal S}$ at $x=n_c$ suggests that this point does not correspond to a Fermi liquid 
even when considering perturbations in electronic hopping $t$.

Then, we consider the lowest order in perturbation expansion,
which is linear in $t$.
Fermi liquids characterize both dilute and dense regimes, but a
singular point separates the LFL from the CFL.

LFL-regime: the first perturbative correction generates an effective model where
the fermionic quasiparticles are $N_{qp}^{LFL}=N_c-N_S$ free electrons moving with  
hopping  energy $t_{ij}$ on the lattice made of the $N-N_S$ non-Kondo sites
(see figure~\ref{Figureschemareseau}~a).
Of course this is valid only if $1-x$ is bigger than the percolation threshold $x_p$.
The only correlation effect in this effective model results from the frozen random
depletion of the Kondo-sites. This picture recovers the early approaches of Nozi\`eres to the single impurity model, which was mapped onto a problem of light conduction electrons scattering on the Kondo singlet~\cite{Nozieres1974,Nozieres1985,Nozieres1998}.
Here, light conduction electrons scatter on the $K-$sites which are effectively excluded.
From CPA approach, we find that the site depletion
leads to an effective conduction band for the $N_{qp}^{LFL}$ quasiparticles, with a renormalized hopping term
$t_{ij}\sqrt{1-x}$.
The local d.o.s is mostly renormalized by the site
depletion, and its total
spectral weight is reduced by a factor $1-x$. The missing weight,  $x$,
is transfered to a higher energy peak, which corresponds to excitations where one of the K-sites has 0 or 2 conduction electrons.
Note that the occupation $N_{qp}^{LFL}$ corresponds to $(n_c-x)/2$ quasiparticles per site and per spin component; this filling is smaller than $1-x$ since $n_c<1$. 

CFL-regime: a perturbation expansion at lowest order provides an effective model where fermionic quasiparticles correspond to the unscreened Kondo impurities. The mapping is very similar to the one described for a periodic Kondo lattice~\cite{Lacroix1985}.
Here, we assume that the $K-$sites concentration exceeds the percolation threshold, i.e., that $x>x_p$.
The unscreened impurities move on a depleted lattice made of the $K-$sites.
There are $N_S-N_c$ unscreened spins moving on $N_S$ sites, and since there motion is
hole-like, the number of quasiparticles is $N_{qp}^{CFL}=2N_S-(N_S-N_c)
=N_S+N_c$. This suggests that the Luttinger theorem applies in the CFL-regime, both electrons and Kondo spins contributing to the Fermi surface.
Here, the correlations have two origins: the random depletion, which also characterizes the LFL-regime, and a supplementary infinite repulsion
which prevents from unphysical double impurity occupation on a same site.
Since the moving holes are Kondo singlets, the lattice hopping is also
renormalized by a factor $1/2$ as explained in Ref.~\cite{Lacroix1985}.
An extra renormalization of the hopping results from the lattice site depletion.
Invoking a CPA approach, the intersite hopping for the correlated quasiparticles is
$t_{ij}\sqrt{x}/2$.
The local d.o.s. is renormalized in two steps: first, the site depletion reduces the bandwidth by a factor $\sqrt{x}$, and the spectral weight by a factor $x$. The missing spectral weight here corresponds to excitations moving an electron from a Kondo singlet to a non Kondo site (with 0 or 2 conduction electrons).
Then, the peak at lowest energy is split again due to the effective Hubbard repulsion. We may interpret the upper forbidden Hubbard band as corresponding to excitations of local Kondo singlets into local triplets. This singlet-triplet excitation mode remains gaped (or pseudo-gaped)
in the opposite limit, $J_K\ll t$, as indicated by mean-field calculations (see
figure~\ref{FiguredosKondoalloy_plusinset}).
Note that the occupation $N_{qp}^{CFL}$ corresponds to $(x+n_c)/2$ quasiparticles per site and per spin component; this filling is smaller than $x/2$ and the associated chemical potential is thus inside the lowest Hubbard band.
This mapping to an effective Hubbard model is exact in the strong Kondo coupling limit, but it is interesting to observe that the numerical results obtained from the mean-field approximation with weak coupling parameters also provide
a splitting of the d.o.s. with a gap opening in the dense regime
(see figure~\ref{FiguredosKondoalloy_plusinset}).

\begin{table}
\begin{tabular}{|l|c|c|}
\hline
~~&LFL-regime&CFL-regime\\
~~&$N_S<N_c$&$N_S>N_c$\\
\hline
Number of quasiparticles&$N_c-N_S$&$N_c+N_S$\\
\hline
Effective intersite hopping&$t_{ij}$&$t_{ij}/2$\\
\hline
Number of effective sites&$N-N_S$&$N_S$\\
\hline
CPA Rescaled hopping&$t_{ij}\sqrt{1-x}$&$t_{ij}\sqrt{x}/2$\\
\hline
Effective Hubbard repulsion&$0$&$\infty$\\
\hline
Spectral weight of the lower band&$1-x$&$x/2$\\
\hline
\end{tabular}
\caption{Characteristics of dense and dilute regimes for large $J_K$. }
\label{Table1}
\end{table}
The characteristics of the strong coupling effective models obtained in the CFL and LFL regimes are compared in 
table~\ref{Table1}. 
The impossibility of connecting analytically these two families of models thus survives to finite $t$ corrections.
We  identify $N_c=N_S$ to a critical point
with the following singularities:
the number of quasiparticles is discontinuous, as well as the number of
effective sites, the effective hopping and the Hubbard repulsion,
which changes from $0$ to $\infty$. For $N_c=N$, this singular point corresponds to a Kondo insulator, and is physically accessible only from the LFL side. For $N_c<N$, the singularity of this point remains, and it is accessible from both LFL and CFL sides.
The critical point is not associated to a symmetry breaking, but to a singular change of the Fermi surface known as a Lifshitz transition (LT)~\cite{Lifshitz1960, Blanter1994}.

\begin{figure}[hhh]
\includegraphics[width=8cm, origin=bc, angle=0]{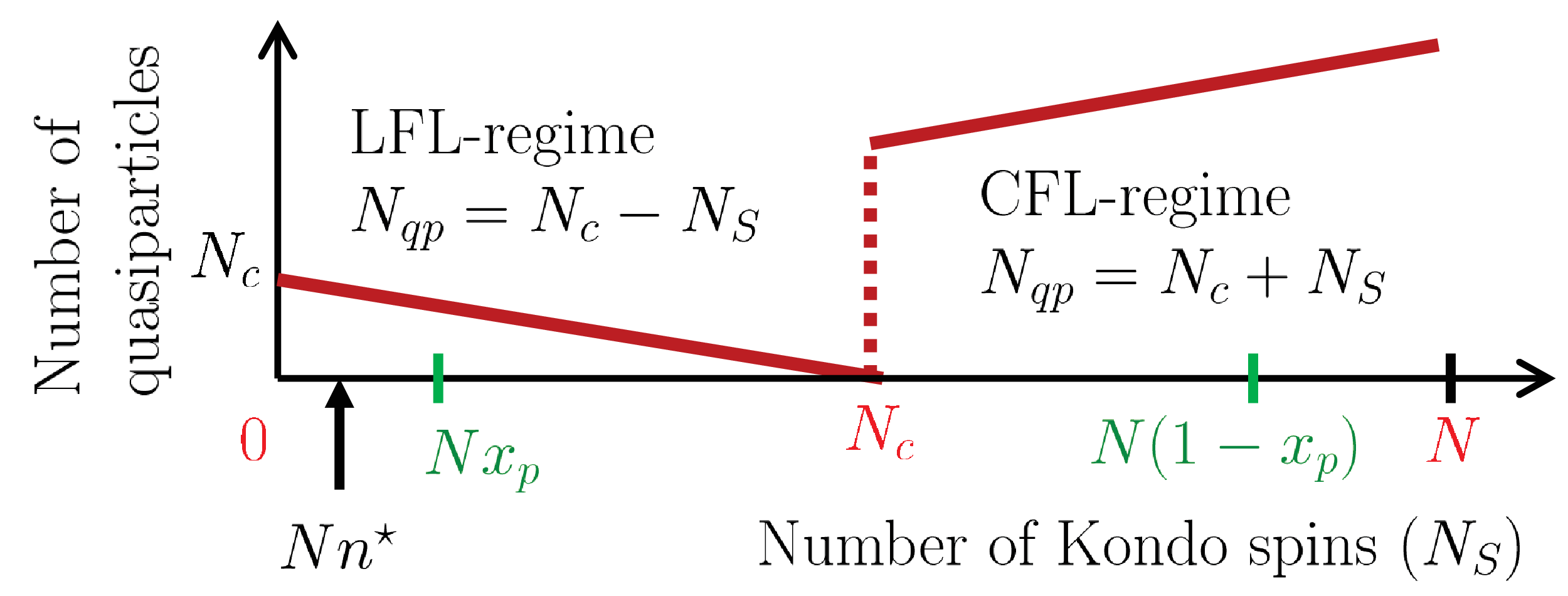}
\caption{Number of quasiparticles as a function of $N_S$ for fixed $N_c$. A transition is expected at $N_S=N_c$, when $x_p<1/2$. 
For realistic (i.e., small) Kondo coupling, only a small part of conduction electrons is involved in the Kondo screening of the LFL-regime, resulting in an intermediate third regime, $Nn^\star <N_S<N_c$. }
\label{FigureQuasiparticleoccupation}
\end{figure}
The number of quasiparticles is ploted in
figure~\ref{FigureQuasiparticleoccupation} as a function of $N_S$ for $J_K\gg t$. The depicted situation corresponds to a KAM which is realized by substitution of non-Kondo (example La) with Kondo (example Ce) atoms, with fixed light electron structure and filling.
In the CFL-regime, the Kondo spins contribute to the volume of the Fermi surface, which is consistent with the general idea of Luttinger theorem.
The formation of this large Fermi surface requires a sufficiently high amount of Kondo impurities, in order to form 
a CFL. But in the LFL-regime, Luttinger theorem is expected to break down.
There, Kondo impurities are too diluted for contributing to the Fermi surface.
Instead, their main effect is rather to capture conduction electrons for the formation of localized Kondo singlets.
Only the resulting $N_c-N_S$ electrons
contribute to the Fermi surface, and scatter on the localized Kondo singlets.
The discontinuity of $N_{qp}$ is related to a LT: the Fermi surface is not defined at the critical point $N_c=N_S$. This is true for $N_c=N_S=1$ because the system is in this case a Kondo insulator, but we predict this situation to occur more generally for
$N_c=N_S$ not necessarily equal to $1$.
Approached from the LFL side, this critical point corresponds to a continuous collapse of the Fermi surface. 
But this collapse is discontinuous from the CFL side. 

Of course, this analysis does not consider possible issues of cluster formations and we also assumed percolation of the depleted effective lattices.
An intermediate Kondo disorder regime with cluster formation is expected for lattices with big percolation threshold, $x_p>1/2$ (which is usual in two-dimensional systems with only nearest neighbor hopping).
The most realistic cases are three-dimensional systems with low percolation threshold $x_p<1/2$ and intermediate electronic filling.
Two situations can thus occur: the LT should be observed for $x_p<n_c<1-x_p$, whilst a Kondo disorder phase
is expected for other values of $n_c$. 
Here, we did not analyze the Kondo disorder regime, where inhomogeneities may control the physical properties as
discussed in~\cite{Dobrosavljevic1992}. We also did not consider the issue of magnetic ordering, but it is expected to be weakened by a factor $x^2$ in a Kondo alloy. Moreover the magnetic interaction, in the strong coupling limit is small ($t^2/J_K$~\cite{Lacroix1985}).

Furthermore, it appeared that Kondo screening in the CFL is a collective effect which involves all electrons and all impurities, whilst in the LFL only a small part 
of conduction electrons, of the order of $T_K/t\equiv n^{\star}$ contribute to the Kondo screening at small coupling~\cite{Burdin2000, Nozieres2005}. This problem does not occur in the strong coupling 
limit but it gives rise to a critical concentration $n^{\star}$ for more realistic coupling. In these cases, the LT at $N_S=N_c$ may become a crossover spread from $x\approx n^{\star}$ to $x=n_c$. It is not clear wether signatures of a Fermi surface could be observed experimentally in this whole crossover regime and the NFL regime may spread as well along a large range of intermediate concentrations separating the LFL from the CFL. 
We propose that in the heavy fermion compound Ce$_x$La$_{1-x}$Ni$_2$Ge$_2$ the system goes from LFL to non Fermi Liquid (NFL) for $x\approx n^{\star}\ll 1$, 
and then from NFL to CFL for $x=n_c\approx 0,6$ (see figure 4 in~\cite{Pikul2012}). 

This mechanism may also provide a possible scenario for the origin of the NFL properties observed recently in 
Ce$_x$La$_{1-x}$PtIn~\cite{Ragel2009}. 
A difference (transition or crossover) between dilute Kondo system and a dense heavy fermion had also been suggested by other theoretical methods, including finite size calculations~\cite{Kaul2007}, quantum Monte Carlo simulations~\cite{Otsuki2010, Watanabe2010a} and Gutzwiller approximation~\cite{Watanabe2010b}.
We propose to analyze more systematically Kondo alloys in order to find other examples of NFL behavior of this type. These studies should be completed by
measurements which analyze the Fermi surface (Angleresolved Photoemission Spectroscopy, quantum oscillations),
in order to test the predicted validity and violation of Luttinger theorem, respectively in the dense and dilute regimes.
One difficulty that may be faced in some compounds would come from a  complex electronic band structure. Usually, several conduction bands are involved, and
the criteria $N_c=N_S$ might be not always reachable. 
The Lifshitz critical point could also have signatures in other experimental realizations of Kondo alloys, in bulk materials or in artificial atom systems like quantum dot arrays or optical lattices.

We acknowledge Philippe Nozi\`eres for innumerable discussions, for carefully reading an earlier version of this manuscript, and for very useful comments.

\bibliography{bibliokondoalloy}

\begin{thebibliography}{50}%
\makeatletter
\providecommand \@ifxundefined [1]{%
 \@ifx{#1\undefined}
}%
\providecommand \@ifnum [1]{%
 \ifnum #1\expandafter \@firstoftwo
 \else \expandafter \@secondoftwo
 \fi
}%
\providecommand \@ifx [1]{%
 \ifx #1\expandafter \@firstoftwo
 \else \expandafter \@secondoftwo
 \fi
}%
\providecommand \natexlab [1]{#1}%
\providecommand \enquote  [1]{``#1''}%
\providecommand \bibnamefont  [1]{#1}%
\providecommand \bibfnamefont [1]{#1}%
\providecommand \citenamefont [1]{#1}%
\providecommand \href@noop [0]{\@secondoftwo}%
\providecommand \href [0]{\begingroup \@sanitize@url \@href}%
\providecommand \@href[1]{\@@startlink{#1}\@@href}%
\providecommand \@@href[1]{\endgroup#1\@@endlink}%
\providecommand \@sanitize@url [0]{\catcode `\\12\catcode `\$12\catcode
  `\&12\catcode `\#12\catcode `\^12\catcode `\_12\catcode `\%12\relax}%
\providecommand \@@startlink[1]{}%
\providecommand \@@endlink[0]{}%
\providecommand \url  [0]{\begingroup\@sanitize@url \@url }%
\providecommand \@url [1]{\endgroup\@href {#1}{\urlprefix }}%
\providecommand \urlprefix  [0]{URL }%
\providecommand \Eprint [0]{\href }%
\providecommand \doibase [0]{http://dx.doi.org/}%
\providecommand \selectlanguage [0]{\@gobble}%
\providecommand \bibinfo  [0]{\@secondoftwo}%
\providecommand \bibfield  [0]{\@secondoftwo}%
\providecommand \translation [1]{[#1]}%
\providecommand \BibitemOpen [0]{}%
\providecommand \bibitemStop [0]{}%
\providecommand \bibitemNoStop [0]{.\EOS\space}%
\providecommand \EOS [0]{\spacefactor3000\relax}%
\providecommand \BibitemShut  [1]{\csname bibitem#1\endcsname}%
\let\auto@bib@innerbib\@empty
\bibitem [{\citenamefont {Hewson}(1993)}]{Hewsonbook1993}%
  \BibitemOpen
  \bibfield  {author} {\bibinfo {author} {\bibfnamefont {A.~C.}\ \bibnamefont
  {Hewson}},\ }\href@noop {} {\emph {\bibinfo {title} {The Kondo Problem to
  Heavy Fermions}}}\ (\bibinfo  {publisher} {Cambridge University Press},\
  \bibinfo {address} {Cambridge, England},\ \bibinfo {year} {1993})\BibitemShut
  {NoStop}%
\bibitem [{\citenamefont {Wilson}(1975)}]{Wilson1975}%
  \BibitemOpen
  \bibfield  {author} {\bibinfo {author} {\bibfnamefont {K.~G.}\ \bibnamefont
  {Wilson}},\ }\href@noop {} {\bibfield  {journal} {\bibinfo  {journal} {Rev.
  Mod. Phys.}\ }\textbf {\bibinfo {volume} {47}},\ \bibinfo {pages} {773}
  (\bibinfo {year} {1975})}\BibitemShut {NoStop}%
\bibitem [{\citenamefont {Bulla}\ \emph {et~al.}(2008)\citenamefont {Bulla},
  \citenamefont {Costi},\ and\ \citenamefont {Pruschke}}]{Bulla2008}%
  \BibitemOpen
  \bibfield  {author} {\bibinfo {author} {\bibfnamefont {R.}~\bibnamefont
  {Bulla}}, \bibinfo {author} {\bibfnamefont {T.~A.}\ \bibnamefont {Costi}}, \
  and\ \bibinfo {author} {\bibfnamefont {T.}~\bibnamefont {Pruschke}},\
  }\href@noop {} {\bibfield  {journal} {\bibinfo  {journal} {Rev. Mod. Phys.}\
  }\textbf {\bibinfo {volume} {80}},\ \bibinfo {pages} {395} (\bibinfo {year}
  {2008})}\BibitemShut {NoStop}%
\bibitem [{\citenamefont {Andrei}(1980)}]{Andrei1980}%
  \BibitemOpen
  \bibfield  {author} {\bibinfo {author} {\bibfnamefont {N.}~\bibnamefont
  {Andrei}},\ }\href@noop {} {\bibfield  {journal} {\bibinfo  {journal} {Phys.
  Rev. Lett.}\ }\textbf {\bibinfo {volume} {45}},\ \bibinfo {pages} {379}
  (\bibinfo {year} {1980})}\BibitemShut {NoStop}%
\bibitem [{\citenamefont {Wiegmann}(1980)}]{Wiegmann1980}%
  \BibitemOpen
  \bibfield  {author} {\bibinfo {author} {\bibfnamefont {P.~B.}\ \bibnamefont
  {Wiegmann}},\ }\href@noop {} {\bibfield  {journal} {\bibinfo  {journal}
  {Physics Letters A}\ }\textbf {\bibinfo {volume} {80}},\ \bibinfo {pages}
  {163} (\bibinfo {year} {1980})}\BibitemShut {NoStop}%
\bibitem [{\citenamefont {Affleck}\ and\ \citenamefont
  {Ludwig}(1991)}]{Affleck1991}%
  \BibitemOpen
  \bibfield  {author} {\bibinfo {author} {\bibfnamefont {I.}~\bibnamefont
  {Affleck}}\ and\ \bibinfo {author} {\bibfnamefont {A.~W.~W.}\ \bibnamefont
  {Ludwig}},\ }\href@noop {} {\bibfield  {journal} {\bibinfo  {journal} {Nucl.
  Phys. B}\ }\textbf {\bibinfo {volume} {352}},\ \bibinfo {pages} {849}
  (\bibinfo {year} {1991})}\BibitemShut {NoStop}%
\bibitem [{\citenamefont {Landau}(1957)}]{Landau1957}%
  \BibitemOpen
  \bibfield  {author} {\bibinfo {author} {\bibfnamefont {L.~D.}\ \bibnamefont
  {Landau}},\ }\href@noop {} {\bibfield  {journal} {\bibinfo  {journal} {JETP}\
  }\textbf {\bibinfo {volume} {3}},\ \bibinfo {pages} {920} (\bibinfo {year}
  {1957})}\BibitemShut {NoStop}%
\bibitem [{\citenamefont {Pines}\ and\ \citenamefont
  {Nozi\`eres}(1966)}]{PinesNozieresbook1966}%
  \BibitemOpen
  \bibfield  {author} {\bibinfo {author} {\bibfnamefont {D.}~\bibnamefont
  {Pines}}\ and\ \bibinfo {author} {\bibfnamefont {P.}~\bibnamefont
  {Nozi\`eres}},\ }\href@noop {} {\emph {\bibinfo {title} {Theory of Quantum
  Liquids Vol. I}}}\ (\bibinfo  {publisher} {W. A. Benjamin},\ \bibinfo
  {address} {New York, USA},\ \bibinfo {year} {1966})\BibitemShut {NoStop}%
\bibitem [{\citenamefont {Nozi\`eres}(1974)}]{Nozieres1974}%
  \BibitemOpen
  \bibfield  {author} {\bibinfo {author} {\bibfnamefont {P.}~\bibnamefont
  {Nozi\`eres}},\ }\href@noop {} {\bibfield  {journal} {\bibinfo  {journal} {J.
  Low Temp. Phys.}\ }\textbf {\bibinfo {volume} {17}},\ \bibinfo {pages} {31}
  (\bibinfo {year} {1974})}\BibitemShut {NoStop}%
\bibitem [{\citenamefont {Doniach}(1977)}]{Doniach1977}%
  \BibitemOpen
  \bibfield  {author} {\bibinfo {author} {\bibfnamefont {S.}~\bibnamefont
  {Doniach}},\ }\href@noop {} {\bibfield  {journal} {\bibinfo  {journal}
  {Physica B\& C}\ }\textbf {\bibinfo {volume} {91}},\ \bibinfo {pages} {231}
  (\bibinfo {year} {1977})}\BibitemShut {NoStop}%
\bibitem [{\citenamefont {Lacroix}\ and\ \citenamefont
  {Cyrot}(1979)}]{Lacroix1979}%
  \BibitemOpen
  \bibfield  {author} {\bibinfo {author} {\bibfnamefont {C.}~\bibnamefont
  {Lacroix}}\ and\ \bibinfo {author} {\bibfnamefont {M.}~\bibnamefont
  {Cyrot}},\ }\href@noop {} {\bibfield  {journal} {\bibinfo  {journal} {Phys.
  Rev. B}\ }\textbf {\bibinfo {volume} {20}},\ \bibinfo {pages} {1969}
  (\bibinfo {year} {1979})}\BibitemShut {NoStop}%
\bibitem [{\citenamefont {Flouquet}(2005)}]{Flouquet2005}%
  \BibitemOpen
  \bibfield  {author} {\bibinfo {author} {\bibfnamefont {J.}~\bibnamefont
  {Flouquet}},\ }\href@noop {} {\emph {\bibinfo {title} {Progress in Low
  Temperature Physics}}}\ (\bibinfo  {publisher} {Elsevier, edited by W.P.
  Halperin},\ \bibinfo {year} {2005})\ pp.\ \bibinfo {pages}
  {139--281}\BibitemShut {NoStop}%
\bibitem [{\citenamefont {Nozi\`eres}(1985)}]{Nozieres1985}%
  \BibitemOpen
  \bibfield  {author} {\bibinfo {author} {\bibfnamefont {P.}~\bibnamefont
  {Nozi\`eres}},\ }\href@noop {} {\bibfield  {journal} {\bibinfo  {journal}
  {Ann. Phys. (Paris)}\ }\textbf {\bibinfo {volume} {10}},\ \bibinfo {pages}
  {19} (\bibinfo {year} {1985})}\BibitemShut {NoStop}%
\bibitem [{\citenamefont {Nozi\`eres}(1998)}]{Nozieres1998}%
  \BibitemOpen
  \bibfield  {author} {\bibinfo {author} {\bibfnamefont {P.}~\bibnamefont
  {Nozi\`eres}},\ }\href@noop {} {\bibfield  {journal} {\bibinfo  {journal}
  {Eur. Phys. J. B}\ }\textbf {\bibinfo {volume} {6}},\ \bibinfo {pages} {447}
  (\bibinfo {year} {1998})}\BibitemShut {NoStop}%
\bibitem [{\citenamefont {Lacroix}(1985)}]{Lacroix1985}%
  \BibitemOpen
  \bibfield  {author} {\bibinfo {author} {\bibfnamefont {C.}~\bibnamefont
  {Lacroix}},\ }\href@noop {} {\bibfield  {journal} {\bibinfo  {journal} {Solid
  State Commun.}\ }\textbf {\bibinfo {volume} {54}},\ \bibinfo {pages} {991}
  (\bibinfo {year} {1985})}\BibitemShut {NoStop}%
\bibitem [{\citenamefont {Lacroix}(1986)}]{Lacroix1986}%
  \BibitemOpen
  \bibfield  {author} {\bibinfo {author} {\bibfnamefont {C.}~\bibnamefont
  {Lacroix}},\ }\href@noop {} {\bibfield  {journal} {\bibinfo  {journal} {J.
  Magn. Magn. Mat.}\ }\textbf {\bibinfo {volume} {60}},\ \bibinfo {pages} {145}
  (\bibinfo {year} {1986})}\BibitemShut {NoStop}%
\bibitem [{\citenamefont {Tahvildar-Zadeh}\ \emph {et~al.}(1997)\citenamefont
  {Tahvildar-Zadeh}, \citenamefont {Jarrell},\ and\ \citenamefont
  {Freericks}}]{Tahvildar1997}%
  \BibitemOpen
  \bibfield  {author} {\bibinfo {author} {\bibfnamefont {A.~N.}\ \bibnamefont
  {Tahvildar-Zadeh}}, \bibinfo {author} {\bibfnamefont {M.}~\bibnamefont
  {Jarrell}}, \ and\ \bibinfo {author} {\bibfnamefont {J.~K.}\ \bibnamefont
  {Freericks}},\ }\href@noop {} {\bibfield  {journal} {\bibinfo  {journal}
  {Phys. Rev. B}\ }\textbf {\bibinfo {volume} {55}} (\bibinfo {year}
  {1997})}\BibitemShut {NoStop}%
\bibitem [{\citenamefont {Tahvildar-Zadeh}\ \emph {et~al.}(1998)\citenamefont
  {Tahvildar-Zadeh}, \citenamefont {Jarrell},\ and\ \citenamefont
  {Freericks}}]{Tahvildar1998}%
  \BibitemOpen
  \bibfield  {author} {\bibinfo {author} {\bibfnamefont {A.~N.}\ \bibnamefont
  {Tahvildar-Zadeh}}, \bibinfo {author} {\bibfnamefont {M.}~\bibnamefont
  {Jarrell}}, \ and\ \bibinfo {author} {\bibfnamefont {J.~K.}\ \bibnamefont
  {Freericks}},\ }\href@noop {} {\bibfield  {journal} {\bibinfo  {journal}
  {Phys. Rev. Lett.}\ }\textbf {\bibinfo {volume} {80}},\ \bibinfo {pages}
  {5168} (\bibinfo {year} {1998})}\BibitemShut {NoStop}%
\bibitem [{\citenamefont {Tahvildar-Zadeh}\ \emph {et~al.}(1999)\citenamefont
  {Tahvildar-Zadeh}, \citenamefont {Jarrell}, \citenamefont {Pruschke},\ and\
  \citenamefont {Freericks}}]{Tahvildar1999}%
  \BibitemOpen
  \bibfield  {author} {\bibinfo {author} {\bibfnamefont {A.~N.}\ \bibnamefont
  {Tahvildar-Zadeh}}, \bibinfo {author} {\bibfnamefont {M.}~\bibnamefont
  {Jarrell}}, \bibinfo {author} {\bibfnamefont {T.}~\bibnamefont {Pruschke}}, \
  and\ \bibinfo {author} {\bibfnamefont {J.~K.}\ \bibnamefont {Freericks}},\
  }\href@noop {} {\bibfield  {journal} {\bibinfo  {journal} {Phys. Rev. B}\
  }\textbf {\bibinfo {volume} {60}},\ \bibinfo {pages} {10782} (\bibinfo {year}
  {1999})}\BibitemShut {NoStop}%
\bibitem [{\citenamefont {Coqblin}\ \emph {et~al.}(2000)\citenamefont
  {Coqblin}, \citenamefont {Gusmao}, \citenamefont {Iglesias}, \citenamefont
  {Lacroix}, \citenamefont {Ruppenthal},\ and\ \citenamefont
  {Simoes}}]{Coqblin2000}%
  \BibitemOpen
  \bibfield  {author} {\bibinfo {author} {\bibfnamefont {B.}~\bibnamefont
  {Coqblin}}, \bibinfo {author} {\bibfnamefont {M.~A.}\ \bibnamefont {Gusmao}},
  \bibinfo {author} {\bibfnamefont {J.~R.}\ \bibnamefont {Iglesias}}, \bibinfo
  {author} {\bibfnamefont {C.}~\bibnamefont {Lacroix}}, \bibinfo {author}
  {\bibfnamefont {A.}~\bibnamefont {Ruppenthal}}, \ and\ \bibinfo {author}
  {\bibfnamefont {A.~S.~D.}\ \bibnamefont {Simoes}},\ }\href@noop {} {\bibfield
   {journal} {\bibinfo  {journal} {Physica B}\ }\textbf {\bibinfo {volume}
  {281}},\ \bibinfo {pages} {50} (\bibinfo {year} {2000})}\BibitemShut
  {NoStop}%
\bibitem [{\citenamefont {Burdin}\ \emph {et~al.}(2000)\citenamefont {Burdin},
  \citenamefont {Georges},\ and\ \citenamefont {Grempel}}]{Burdin2000}%
  \BibitemOpen
  \bibfield  {author} {\bibinfo {author} {\bibfnamefont {S.}~\bibnamefont
  {Burdin}}, \bibinfo {author} {\bibfnamefont {A.}~\bibnamefont {Georges}}, \
  and\ \bibinfo {author} {\bibfnamefont {D.~R.}\ \bibnamefont {Grempel}},\
  }\href@noop {} {\bibfield  {journal} {\bibinfo  {journal} {Phys. Rev. Lett.}\
  }\textbf {\bibinfo {volume} {85}},\ \bibinfo {pages} {1048} (\bibinfo {year}
  {2000})}\BibitemShut {NoStop}%
\bibitem [{\citenamefont {Costi}\ and\ \citenamefont
  {Manini}(2002)}]{Costi2002}%
  \BibitemOpen
  \bibfield  {author} {\bibinfo {author} {\bibfnamefont {T.~A.}\ \bibnamefont
  {Costi}}\ and\ \bibinfo {author} {\bibfnamefont {N.}~\bibnamefont {Manini}},\
  }\href@noop {} {\bibfield  {journal} {\bibinfo  {journal} {J. Low Temp.
  Phys.}\ }\textbf {\bibinfo {volume} {126}},\ \bibinfo {pages} {835} (\bibinfo
  {year} {2002})}\BibitemShut {NoStop}%
\bibitem [{\citenamefont {Coqblin}\ \emph {et~al.}(2003)\citenamefont
  {Coqblin}, \citenamefont {Lacroix}, \citenamefont {Gusmao},\ and\
  \citenamefont {Iglesias}}]{Coqblin2003}%
  \BibitemOpen
  \bibfield  {author} {\bibinfo {author} {\bibfnamefont {B.}~\bibnamefont
  {Coqblin}}, \bibinfo {author} {\bibfnamefont {C.}~\bibnamefont {Lacroix}},
  \bibinfo {author} {\bibfnamefont {M.~A.}\ \bibnamefont {Gusmao}}, \ and\
  \bibinfo {author} {\bibfnamefont {J.~R.}\ \bibnamefont {Iglesias}},\
  }\href@noop {} {\bibfield  {journal} {\bibinfo  {journal} {Phys. Rev. B}\
  }\textbf {\bibinfo {volume} {67}},\ \bibinfo {pages} {064417} (\bibinfo
  {year} {2003})}\BibitemShut {NoStop}%
\bibitem [{\citenamefont {Nozi\`eres}(2005)}]{Nozieres2005}%
  \BibitemOpen
  \bibfield  {author} {\bibinfo {author} {\bibfnamefont {P.}~\bibnamefont
  {Nozi\`eres}},\ }\href@noop {} {\bibfield  {journal} {\bibinfo  {journal} {J.
  Phys. Soc. Jpn.}\ }\textbf {\bibinfo {volume} {74}},\ \bibinfo {pages} {4}
  (\bibinfo {year} {2005})}\BibitemShut {NoStop}%
\bibitem [{\citenamefont {Burdin}\ and\ \citenamefont
  {Fulde}(2007)}]{Burdin2007}%
  \BibitemOpen
  \bibfield  {author} {\bibinfo {author} {\bibfnamefont {S.}~\bibnamefont
  {Burdin}}\ and\ \bibinfo {author} {\bibfnamefont {P.}~\bibnamefont {Fulde}},\
  }\href@noop {} {\bibfield  {journal} {\bibinfo  {journal} {Phys. Rev. B}\
  }\textbf {\bibinfo {volume} {76}},\ \bibinfo {pages} {104425} (\bibinfo
  {year} {2007})}\BibitemShut {NoStop}%
\bibitem [{\citenamefont {Burdin}\ and\ \citenamefont
  {Zlati\`c}(2009)}]{Burdin2009}%
  \BibitemOpen
  \bibfield  {author} {\bibinfo {author} {\bibfnamefont {S.}~\bibnamefont
  {Burdin}}\ and\ \bibinfo {author} {\bibfnamefont {V.}~\bibnamefont
  {Zlati\`c}},\ }\href@noop {} {\bibfield  {journal} {\bibinfo  {journal}
  {Phys. Rev. B}\ }\textbf {\bibinfo {volume} {79}},\ \bibinfo {pages} {115139}
  (\bibinfo {year} {2009})}\BibitemShut {NoStop}%
\bibitem [{\citenamefont {Oitmaa}\ and\ \citenamefont
  {Zheng}(2003)}]{Oitmaa2003}%
  \BibitemOpen
  \bibfield  {author} {\bibinfo {author} {\bibfnamefont {J.}~\bibnamefont
  {Oitmaa}}\ and\ \bibinfo {author} {\bibfnamefont {W.}~\bibnamefont {Zheng}},\
  }\href@noop {} {\bibfield  {journal} {\bibinfo  {journal} {Phys. Rev. B}\
  }\textbf {\bibinfo {volume} {67}},\ \bibinfo {pages} {214407} (\bibinfo
  {year} {2003})}\BibitemShut {NoStop}%
\bibitem [{\citenamefont {Kim}\ and\ \citenamefont {Kim}(2007)}]{Kim2007}%
  \BibitemOpen
  \bibfield  {author} {\bibinfo {author} {\bibfnamefont {K.~S.}\ \bibnamefont
  {Kim}}\ and\ \bibinfo {author} {\bibfnamefont {M.~D.}\ \bibnamefont {Kim}},\
  }\href@noop {} {\bibfield  {journal} {\bibinfo  {journal} {Phys. Rev. B}\
  }\textbf {\bibinfo {volume} {75}},\ \bibinfo {pages} {035117} (\bibinfo
  {year} {2007})}\BibitemShut {NoStop}%
\bibitem [{\citenamefont {Kaul}\ and\ \citenamefont {Vojta}(2007)}]{Kaul2007}%
  \BibitemOpen
  \bibfield  {author} {\bibinfo {author} {\bibfnamefont {R.~K.}\ \bibnamefont
  {Kaul}}\ and\ \bibinfo {author} {\bibfnamefont {M.}~\bibnamefont {Vojta}},\
  }\href@noop {} {\bibfield  {journal} {\bibinfo  {journal} {Phys. Rev. B}\
  }\textbf {\bibinfo {volume} {75}},\ \bibinfo {pages} {132407} (\bibinfo
  {year} {2007})}\BibitemShut {NoStop}%
\bibitem [{\citenamefont {Anderson}(1970)}]{Anderson1970}%
  \BibitemOpen
  \bibfield  {author} {\bibinfo {author} {\bibfnamefont {P.~W.}\ \bibnamefont
  {Anderson}},\ }\href@noop {} {\bibfield  {journal} {\bibinfo  {journal} {J.
  Phys. C: Solid State Phys.}\ }\textbf {\bibinfo {volume} {3}},\ \bibinfo
  {pages} {2436} (\bibinfo {year} {1970})}\BibitemShut {NoStop}%
\bibitem [{\citenamefont {Abrikosov}\ \emph {et~al.}(1963)\citenamefont
  {Abrikosov}, \citenamefont {Gorkov},\ and\ \citenamefont
  {Dzyloshinski}}]{Abrikosovbook1963}%
  \BibitemOpen
  \bibfield  {author} {\bibinfo {author} {\bibfnamefont {A.~A.}\ \bibnamefont
  {Abrikosov}}, \bibinfo {author} {\bibfnamefont {L.~P.}\ \bibnamefont
  {Gorkov}}, \ and\ \bibinfo {author} {\bibfnamefont {I.~E.}\ \bibnamefont
  {Dzyloshinski}},\ }\href@noop {} {\emph {\bibinfo {title} {Methods of Quantum
  Field Theory in Statistical Physics}}}\ (\bibinfo  {publisher} {Dover
  Publications Inc},\ \bibinfo {address} {New York, USA},\ \bibinfo {year}
  {1963})\BibitemShut {NoStop}%
\bibitem [{\citenamefont {Coleman}(1983)}]{Coleman1983}%
  \BibitemOpen
  \bibfield  {author} {\bibinfo {author} {\bibfnamefont {P.}~\bibnamefont
  {Coleman}},\ }\href@noop {} {\bibfield  {journal} {\bibinfo  {journal} {Phys.
  Rev. B}\ }\textbf {\bibinfo {volume} {28}},\ \bibinfo {pages} {28} (\bibinfo
  {year} {1983})}\BibitemShut {NoStop}%
\bibitem [{\citenamefont {Read}\ \emph {et~al.}(1984)\citenamefont {Read},
  \citenamefont {Newns},\ and\ \citenamefont {Doniach}}]{Read1984}%
  \BibitemOpen
  \bibfield  {author} {\bibinfo {author} {\bibfnamefont {N.}~\bibnamefont
  {Read}}, \bibinfo {author} {\bibfnamefont {D.~M.}\ \bibnamefont {Newns}}, \
  and\ \bibinfo {author} {\bibfnamefont {S.}~\bibnamefont {Doniach}},\
  }\href@noop {} {\bibfield  {journal} {\bibinfo  {journal} {Phys. Rev. B}\
  }\textbf {\bibinfo {volume} {30}},\ \bibinfo {pages} {3841} (\bibinfo {year}
  {1984})}\BibitemShut {NoStop}%
\bibitem [{\citenamefont {George}\ \emph {et~al.}(1996)\citenamefont {George},
  \citenamefont {Kotliar}, \citenamefont {Krauth},\ and\ \citenamefont
  {Rozenberg}}]{George1996}%
  \BibitemOpen
  \bibfield  {author} {\bibinfo {author} {\bibfnamefont {A.}~\bibnamefont
  {George}}, \bibinfo {author} {\bibfnamefont {G.}~\bibnamefont {Kotliar}},
  \bibinfo {author} {\bibfnamefont {W.}~\bibnamefont {Krauth}}, \ and\ \bibinfo
  {author} {\bibfnamefont {M.~J.}\ \bibnamefont {Rozenberg}},\ }\href@noop {}
  {\bibfield  {journal} {\bibinfo  {journal} {Rev. Mod. Phys.}\ }\textbf
  {\bibinfo {volume} {68}},\ \bibinfo {pages} {13} (\bibinfo {year}
  {1996})}\BibitemShut {NoStop}%
\bibitem [{\citenamefont {Metzner}\ and\ \citenamefont
  {Vollhardt}(1989)}]{Metzner1989}%
  \BibitemOpen
  \bibfield  {author} {\bibinfo {author} {\bibfnamefont {W.}~\bibnamefont
  {Metzner}}\ and\ \bibinfo {author} {\bibfnamefont {D.}~\bibnamefont
  {Vollhardt}},\ }\href@noop {} {\bibfield  {journal} {\bibinfo  {journal}
  {Phys. Rev. Lett.}\ }\textbf {\bibinfo {volume} {62}},\ \bibinfo {pages}
  {324} (\bibinfo {year} {1989})}\BibitemShut {NoStop}%
\bibitem [{\citenamefont {Blackman}\ \emph {et~al.}(1971)\citenamefont
  {Blackman}, \citenamefont {Esterling},\ and\ \citenamefont
  {Berk}}]{Blackman1971}%
  \BibitemOpen
  \bibfield  {author} {\bibinfo {author} {\bibfnamefont {J.~A.}\ \bibnamefont
  {Blackman}}, \bibinfo {author} {\bibfnamefont {D.~M.}\ \bibnamefont
  {Esterling}}, \ and\ \bibinfo {author} {\bibfnamefont {N.~F.}\ \bibnamefont
  {Berk}},\ }\href@noop {} {\bibfield  {journal} {\bibinfo  {journal} {Phys.
  Rev. B}\ }\textbf {\bibinfo {volume} {4}},\ \bibinfo {pages} {2412} (\bibinfo
  {year} {1971})}\BibitemShut {NoStop}%
\bibitem [{\citenamefont {Esterling}(1975)}]{Esterling1975}%
  \BibitemOpen
  \bibfield  {author} {\bibinfo {author} {\bibfnamefont {D.~M.}\ \bibnamefont
  {Esterling}},\ }\href@noop {} {\bibfield  {journal} {\bibinfo  {journal}
  {Phys. Rev. B}\ }\textbf {\bibinfo {volume} {12}},\ \bibinfo {pages} {1596}
  (\bibinfo {year} {1975})}\BibitemShut {NoStop}%
\bibitem [{\citenamefont {Hoshino}\ and\ \citenamefont
  {Kurata}(1979)}]{Hoshino1979}%
  \BibitemOpen
  \bibfield  {author} {\bibinfo {author} {\bibfnamefont {K.}~\bibnamefont
  {Hoshino}}\ and\ \bibinfo {author} {\bibfnamefont {Y.}~\bibnamefont
  {Kurata}},\ }\href@noop {} {\bibfield  {journal} {\bibinfo  {journal} {J.
  Phys. F: Metal Phys.}\ }\textbf {\bibinfo {volume} {9}},\ \bibinfo {pages}
  {131} (\bibinfo {year} {1979})}\BibitemShut {NoStop}%
\bibitem [{\citenamefont {Kurata}(1979)}]{Kurata1979}%
  \BibitemOpen
  \bibfield  {author} {\bibinfo {author} {\bibfnamefont {Y.}~\bibnamefont
  {Kurata}},\ }\href@noop {} {\bibfield  {journal} {\bibinfo  {journal} {J.
  Phys. F: Metal Phys.}\ }\textbf {\bibinfo {volume} {10}},\ \bibinfo {pages}
  {893} (\bibinfo {year} {1979})}\BibitemShut {NoStop}%
\bibitem [{\citenamefont {Li}\ \emph {et~al.}(1994)\citenamefont {Li},
  \citenamefont {Chen},\ and\ \citenamefont {Xiao}}]{Li1994}%
  \BibitemOpen
  \bibfield  {author} {\bibinfo {author} {\bibfnamefont {Z.~Z.}\ \bibnamefont
  {Li}}, \bibinfo {author} {\bibfnamefont {W.~X.~C.}\ \bibnamefont {Chen}}, \
  and\ \bibinfo {author} {\bibfnamefont {M.~W.}\ \bibnamefont {Xiao}},\
  }\href@noop {} {\bibfield  {journal} {\bibinfo  {journal} {Phys. Rev. B}\
  }\textbf {\bibinfo {volume} {50}},\ \bibinfo {pages} {11332} (\bibinfo {year}
  {1994})}\BibitemShut {NoStop}%
\bibitem [{\citenamefont {Grenzebach}\ \emph {et~al.}(2008)\citenamefont
  {Grenzebach}, \citenamefont {Anders}, \citenamefont {Czycholl},\ and\
  \citenamefont {Pruschke}}]{Grenzebach2008}%
  \BibitemOpen
  \bibfield  {author} {\bibinfo {author} {\bibfnamefont {C.}~\bibnamefont
  {Grenzebach}}, \bibinfo {author} {\bibfnamefont {F.}~\bibnamefont {Anders}},
  \bibinfo {author} {\bibfnamefont {G.}~\bibnamefont {Czycholl}}, \ and\
  \bibinfo {author} {\bibfnamefont {T.}~\bibnamefont {Pruschke}},\ }\href@noop
  {} {\bibfield  {journal} {\bibinfo  {journal} {Phys. Rev. B}\ }\textbf
  {\bibinfo {volume} {77}},\ \bibinfo {pages} {115125} (\bibinfo {year}
  {2008})}\BibitemShut {NoStop}%
\bibitem [{\citenamefont {Zhang}(2002)}]{Zhang2002}%
  \BibitemOpen
  \bibfield  {author} {\bibinfo {author} {\bibfnamefont {S.}~\bibnamefont
  {Zhang}},\ }\href@noop {} {\bibfield  {journal} {\bibinfo  {journal} {Phys.
  Rev. B}\ }\textbf {\bibinfo {volume} {65}},\ \bibinfo {pages} {064407}
  (\bibinfo {year} {2002})}\BibitemShut {NoStop}%
\bibitem [{\citenamefont {Lifshitz}(1960)}]{Lifshitz1960}%
  \BibitemOpen
  \bibfield  {author} {\bibinfo {author} {\bibfnamefont {I.~M.}\ \bibnamefont
  {Lifshitz}},\ }\href@noop {} {\bibfield  {journal} {\bibinfo  {journal} {Sov.
  Phys. JETP}\ }\textbf {\bibinfo {volume} {11}},\ \bibinfo {pages} {1130}
  (\bibinfo {year} {1960})}\BibitemShut {NoStop}%
\bibitem [{\citenamefont {Blanter}\ \emph {et~al.}(1994)\citenamefont
  {Blanter}, \citenamefont {Kaganov}, \citenamefont {Pantsulaya},\ and\
  \citenamefont {Varlamov}}]{Blanter1994}%
  \BibitemOpen
  \bibfield  {author} {\bibinfo {author} {\bibfnamefont {Y.}~\bibnamefont
  {Blanter}}, \bibinfo {author} {\bibfnamefont {M.}~\bibnamefont {Kaganov}},
  \bibinfo {author} {\bibfnamefont {A.}~\bibnamefont {Pantsulaya}}, \ and\
  \bibinfo {author} {\bibfnamefont {A.}~\bibnamefont {Varlamov}},\ }\href
  {\doibase 10.1016/0370-1573(94)90103-1} {\bibfield  {journal} {\bibinfo
  {journal} {Physics Reports}\ }\textbf {\bibinfo {volume} {245}},\ \bibinfo
  {pages} {159} (\bibinfo {year} {1994})}\BibitemShut {NoStop}%
\bibitem [{\citenamefont {Dobrosavljevi\'c}\ \emph {et~al.}(1992)\citenamefont
  {Dobrosavljevi\'c}, \citenamefont {Kirkpatrick},\ and\ \citenamefont
  {Kotliar}}]{Dobrosavljevic1992}%
  \BibitemOpen
  \bibfield  {author} {\bibinfo {author} {\bibfnamefont {V.}~\bibnamefont
  {Dobrosavljevi\'c}}, \bibinfo {author} {\bibfnamefont {T.}~\bibnamefont
  {Kirkpatrick}}, \ and\ \bibinfo {author} {\bibfnamefont {G.}~\bibnamefont
  {Kotliar}},\ }\href@noop {} {\bibfield  {journal} {\bibinfo  {journal} {Phys.
  Rev. Lett.}\ }\textbf {\bibinfo {volume} {69}},\ \bibinfo {pages} {1113}
  (\bibinfo {year} {1992})}\BibitemShut {NoStop}%
\bibitem [{\citenamefont {Pikul}\ \emph {et~al.}(2012)\citenamefont {Pikul},
  \citenamefont {Stockert}, \citenamefont {Steppke}, \citenamefont {Cichorek},
  \citenamefont {Hartmann}, \citenamefont {Caroca-Canales}, \citenamefont
  {Oeschler}, \citenamefont {Brando}, \citenamefont {Geibel},\ and\
  \citenamefont {Steglich}}]{Pikul2012}%
  \BibitemOpen
  \bibfield  {author} {\bibinfo {author} {\bibfnamefont {A.~P.}\ \bibnamefont
  {Pikul}}, \bibinfo {author} {\bibfnamefont {U.}~\bibnamefont {Stockert}},
  \bibinfo {author} {\bibfnamefont {A.}~\bibnamefont {Steppke}}, \bibinfo
  {author} {\bibfnamefont {T.}~\bibnamefont {Cichorek}}, \bibinfo {author}
  {\bibfnamefont {S.}~\bibnamefont {Hartmann}}, \bibinfo {author}
  {\bibfnamefont {N.}~\bibnamefont {Caroca-Canales}}, \bibinfo {author}
  {\bibfnamefont {N.}~\bibnamefont {Oeschler}}, \bibinfo {author}
  {\bibfnamefont {M.}~\bibnamefont {Brando}}, \bibinfo {author} {\bibfnamefont
  {C.}~\bibnamefont {Geibel}}, \ and\ \bibinfo {author} {\bibfnamefont
  {F.}~\bibnamefont {Steglich}},\ }\href@noop {} {\bibfield  {journal}
  {\bibinfo  {journal} {Phys. Rev. Lett.}\ }\textbf {\bibinfo {volume} {108}},\
  \bibinfo {pages} {066405} (\bibinfo {year} {2012})}\BibitemShut {NoStop}%
\bibitem [{\citenamefont {Ragel}\ \emph {et~al.}(2009)\citenamefont {Ragel},
  \citenamefont {de~V.~du Plessis},\ and\ \citenamefont {Strydom}}]{Ragel2009}%
  \BibitemOpen
  \bibfield  {author} {\bibinfo {author} {\bibfnamefont {F.~C.}\ \bibnamefont
  {Ragel}}, \bibinfo {author} {\bibfnamefont {P.}~\bibnamefont {de~V.~du
  Plessis}}, \ and\ \bibinfo {author} {\bibfnamefont {A.~M.}\ \bibnamefont
  {Strydom}},\ }\href@noop {} {\bibfield  {journal} {\bibinfo  {journal} {J.
  Phys. Cond. Matt.}\ }\textbf {\bibinfo {volume} {21}},\ \bibinfo {pages}
  {046008} (\bibinfo {year} {2009})}\BibitemShut {NoStop}%
\bibitem [{\citenamefont {Otsuki}\ \emph {et~al.}(2010)\citenamefont {Otsuki},
  \citenamefont {Kusunose},\ and\ \citenamefont {Kuramoto}}]{Otsuki2010}%
  \BibitemOpen
  \bibfield  {author} {\bibinfo {author} {\bibfnamefont {J.}~\bibnamefont
  {Otsuki}}, \bibinfo {author} {\bibfnamefont {H.}~\bibnamefont {Kusunose}}, \
  and\ \bibinfo {author} {\bibfnamefont {Y.}~\bibnamefont {Kuramoto}},\
  }\href@noop {} {\bibfield  {journal} {\bibinfo  {journal} {J. Phys. Soc.
  Jpn.}\ }\textbf {\bibinfo {volume} {79}},\ \bibinfo {pages} {114709}
  (\bibinfo {year} {2010})}\BibitemShut {NoStop}%
\bibitem [{\citenamefont {Watanabe}\ and\ \citenamefont
  {Ogata}(2010{\natexlab{a}})}]{Watanabe2010a}%
  \BibitemOpen
  \bibfield  {author} {\bibinfo {author} {\bibfnamefont {H.}~\bibnamefont
  {Watanabe}}\ and\ \bibinfo {author} {\bibfnamefont {M.}~\bibnamefont
  {Ogata}},\ }\href@noop {} {\bibfield  {journal} {\bibinfo  {journal} {J.
  Phys.:Conf. Ser.}\ }\textbf {\bibinfo {volume} {200}},\ \bibinfo {pages}
  {012221} (\bibinfo {year} {2010}{\natexlab{a}})}\BibitemShut {NoStop}%
\bibitem [{\citenamefont {Watanabe}\ and\ \citenamefont
  {Ogata}(2010{\natexlab{b}})}]{Watanabe2010b}%
  \BibitemOpen
  \bibfield  {author} {\bibinfo {author} {\bibfnamefont {H.}~\bibnamefont
  {Watanabe}}\ and\ \bibinfo {author} {\bibfnamefont {M.}~\bibnamefont
  {Ogata}},\ }\href@noop {} {\bibfield  {journal} {\bibinfo  {journal} {Phys.
  Rev. B}\ }\textbf {\bibinfo {volume} {81}},\ \bibinfo {pages} {113111}
  (\bibinfo {year} {2010}{\natexlab{b}})}\BibitemShut {NoStop}%
\end{thebibliography}%

\end{document}